# Conservation Laws and Stability of Field Theories of Derived Type

Dmitry S. Kaparulin

Faculty of Physics, Tomsk State University, Tomsk 634050, Russia;
Correspondence: dsc@phys.tsu.ru

**Abstract:** We consider the issue of correspondence between symmetries and conserved quantities in the class of linear relativistic higher-derivative theories of derived type. In this class of models the wave operator is a polynomial in another formally self-adjoint operator, while each isometry of space-time gives rise to the series of symmetries of action functional. If the wave operator is given by $n$-th-order polynomial then this series includes $n$ independent entries, which can be explicitly constructed. The Noether theorem is then used to construct an $n$-parameter set of second-rank conserved tensors. The canonical energy-momentum tensor is included in the series, while the other entries define independent integrals of motion. The Lagrange anchor concept is applied to connect the general conserved tensor in the series with the original space-time translation symmetry. This result is interpreted as existence of multiple energy-momentum tensors in the class of derived systems. To study stability we seek for bounded-conserved quantities that are connected with the time translations. We observe that the derived theory is stable if its wave operator is defined by a polynomial with simple and real roots. The general constructions are illustrated by the examples of the Pais–Uhlenbeck oscillator, higher-derivative scalar field, and extended Chern–Simons theory.

## 1. Introduction

Once the Noether theorem [1] is applied to the higher-derivative theories, the models whose Lagrangians involve second and higher-derivatives in time, the canonical energy usually appears to be unbounded. This is often interpreted as instability of higher-derivative dynamics [2]. The presence of classical trajectories with runaway behavior and the absence of a well-defined vacuum state with the lowest energy at the quantum level are considered as typical indicators of stability problems. The recent research [3–5] demonstrates that the models with unbounded classical energy are not necessarily unstable. Various ideas were applied to the study of stability of higher-derivative theories, including the non-Hermitian quantum mechanics [6–8], alternative Hamiltonian formulations [9–12], adiabatic invariants [13], and special boundary conditions [14]. For constrained systems, the energy can be bounded on-shell due to constraints [15–17]. The f(R)-gravity [18–20] is the most studied model of such a type.

In [21], it has been observed that the higher-derivative dynamics can be stabilized by another bounded-conserved quantity. Even though the Noether theorem associates this bounded quantity with a certain higher symmetry, the Lagrange anchor can be used to connect the additional integral of motion with the time translation. The Lagrange anchor was first introduced to quantize non-variational models in [22]. Later, it was observed that it also connects symmetries and conserved quantities in both Lagrangian and non-Lagrangian theories [23]. In the first-order formalism, the Lagrange anchor defines the Poisson bracket [24]. The bounded-conserved quantity, which is connected with the time translation symmetry, serves as Hamiltonian with respect to this Poisson bracket. In this way, the stability of dynamics is retained at both classical and quantum levels.

In [25], an interesting class of higher-derivative theories of derived type was introduced. In these models the wave operator is a constant coefficient polynomial (characteristic polynomial) in another formally self-adjoint operator of lower order. The Pais–Uhlenbeck theory [26], Podolsky electrodynamics [27], conformal gravities in four and six dimensions [28,29], and extended Chern–Simons [30] are particular examples of theories of derived type. The general result of [25] is the following: the derived theory with the *n*-th-order characteristic polynomial admits *n*-parameter series of conserved quantities, which can be bounded or unbounded. The bounded-conserved quantities can stabilize the classical dynamics of the theory. It was also claimed (without detailed analysis) that all of the conserved quantities in the series are connected with the time translation symmetry by a Lagrange anchor.

In the present article, we study the stability of higher-derivative dynamics from the viewpoint of a more general correspondence between symmetries and conservation laws, which is established by the Lagrange anchor. We show that the derived model with the *n*-th-order characteristic polynomial admits an *n*-parameter series of Lagrange anchors, which connects the general conserved quantity with time translation symmetry. To get this result we reformulate the correspondence between symmetries and conserved quantities in terms of algebra of polynomials and apply the Bezout Lemma. To study stability we address the issue of correspondence between the time translations and bounded-conserved quantities. This problem is special because the bounded-conserved quantities are not general representatives of the conserved quantity series. We conclude that in non-singular theories a bounded-conserved quantity can be connected with the time translation if all of the roots of the characteristic polynomial are real and simple. As for gauge models, a more accurate analysis is performed.

The rest of the paper is organized as follows. In Section 2, we recall some basic facts about symmetries, characteristics, and conservation laws in linear systems. In Section 3, we introduce the Lagrange anchor and establish a correspondence between symmetries and conservation laws. The class of derived models is introduced in Section 4. Section 5 studies the issue of the relationship between the bounded quantities and time translations. We also obtain the stability conditions of higher-derivative theories of derived type in Section 5. Section 6 illustrates general constructions in the theories of the Pais–Uhlenbeck oscillator, higher-derivative scalar, and extended Chern–Simons. The conclusion summarizes the results.

## 2. Symmetries, Characteristics, and Conserved Quantities of Linear Systems

Given the *d*-dimensional Minkowski space[1] with the local coordinates $x^\mu$, $\mu = 0,1,...,d-1$, we consider the set of fields $\varphi^i(x)$. The multi-index *i* includes all the tensor, spinor, and isotopic indices, which label the fields. We assume the existence of an appropriate constant metric which can be used to raise and lower the multi-indices. This gives rise to the inner product of fields,

$$\langle \varphi, \psi \rangle = \varphi_i(x)\psi^i(x) \tag{1}$$

(no integration over space-time). Zero boundary conditions are assumed for the fields at infinity. In this setting, the most general linear theory reads as PDE,

$$T_i(\varphi) = M_{ij}(\partial)\varphi^j = 0, \tag{2}$$

where summation over repeated index is implied, and *M* is the matrix differential operator. By the matrix differential operator, we mean the matrix whose entries are polynomials in the formal variable $\partial_\mu = \partial/\partial x^\mu$.

The formal adjoint of the matrix differential operator *M* is defined as follows:

$$M^\dagger{}_{ij}(\partial) = M_{ji}(-\partial). \tag{3}$$

If the wave operator is formally self-adjoint, the corresponding equations (1) are variational with the action functional

---

[1] We use the mostly minus convention for the space-time metric throughout this paper.

$$S[\varphi(x)] = \frac{1}{2}\int \langle \varphi, M\varphi \rangle d^d x. \tag{4}$$

In this class of models, $M$ is often called the wave operator. The following identity relates the operator and its adjoint:

$$\frac{1}{2}\left(\langle \varphi, M\psi \rangle - \langle M^\dagger \varphi, \psi \rangle\right) = \partial_\mu \, (j_M(\varphi,\psi))^\mu. \tag{5}$$

Here, $\psi$ and $\varphi$ are test functions, and the right-hand side is a divergence of some vector. The index $M$ in $j_M$ labels the operator $M$, which is involved in the left-hand side of equation (5).

We allow gauge freedom for the model (2). If the wave operator $M$ has right null-vectors in the class of matrix differential operators, i.e.,

$$M(\partial)R_\alpha(\partial) \equiv 0, \tag{6}$$

with $\alpha$ being some multi-index, equations (2) are invariant with respect to the gauge transformation

$$\delta_\varepsilon \varphi^i = R^i_\alpha(\partial)\varepsilon^\alpha. \tag{7}$$

Here, $\varepsilon^\alpha = \varepsilon^\alpha(x)$ are functions of space-time coordinates, and summation over the repeated index $\alpha$ is assumed. The matrix differential operators $R_\alpha$ are called gauge generators.

The wave operator can have left null-vectors in the class of matrix differential operators. All such null-vectors determine gauge identities between equations of motion of the following form:

$$Z_A(\partial)T(\varphi) \equiv 0, \tag{8}$$

where the multi-index $A$ labels gauge identities. For linear variational theories, the identity generators are adjoint of gauge generators,

$$Z_A(\partial) = R^\dagger_\alpha(\partial), \quad A \equiv \alpha. \tag{9}$$

The statement about the relation between gauge symmetries and gauge identities is often called the second Noether theorem, and equality (9) is a particular form of it. In non-variational theories, the gauge symmetries and identity generators are unrelated to each other.

The matrix differential operator $L$ is called the symmetry[2] of linear theory (2) if it is interchangeable with $M$ in the following sense:

$$M(\partial)L(\partial) = Q(\partial)M(\partial), \tag{10}$$

with $Q$ being some matrix differential operator. Symmetry induces a linear transformation of the fields that preserves the mass shell (2):

$$\delta_\xi \varphi = \xi L(\partial)\varphi, \quad \delta_\xi T(\varphi) = \xi L(\partial)T(\varphi) \approx 0. \tag{11}$$

Here, the constant $\xi$ is the infinitesimal transformation parameter, and the sign $\approx$ means equality modulo equations (2). The symmetries of linear theory form an associative algebra with respect to the composition of the operators [31]. Equation (10) has a lot of trivial solutions of the form

$$L_{\text{triv}}(\partial) = U(\partial)M(\partial) + R_\alpha(\partial)U^\alpha(\partial), \tag{12}$$

where $U$ and $U^\alpha$ are some matrix differential operators. These symmetries are present in every theory and do not contain any valuable information about the dynamics of the model. In what follows, we systematically ignore them.

---

[2] The notion of symmetry can be introduced in various ways. In this article we give non-rigorous explanations about symmetries, which are sufficient for our consideration. For systematical introduction into the subject we refer to the book [31].

We term the symmetry variational if the transformation (11) preserves the action functional (4). The defining condition for the variational symmetry reads

$$L^\dagger(\partial)M(\partial) + M(\partial)L(\partial) = 0, \tag{13}$$

which is relation (10) for $Q = L^\dagger$. Trivial variational symmetries have the form

$$L^{\text{var}}_{\text{triv}}(\partial) = G(\partial)M(\partial) + R_\alpha(\partial)U^\alpha(\partial), \quad G^\dagger = -G. \tag{14}$$

The vector-valued function of the fields and their derivatives *j* is called a conserved current of the model (4) if its divergence vanishes on the mass shell, i.e.,

$$\partial_\mu j^\mu(\varphi, \partial\varphi, \partial^2\varphi, \ldots) = \langle N(\varphi), T(\varphi) \rangle \tag{15}$$

for some characteristics, *N* being the function of fields and its derivatives. The conserved current *j* defines the integral of motion:

$$J = \int j^0(\varphi)d^{d-1}x, \tag{16}$$

where the integration is held over the space coordinates. By construction, *J* is a constant for all the solutions to the classical equations (2).

The above definition of conserved current and characteristic has natural ambiguities. Two conserved currents are considered equivalent if their difference is the divergence of some anti-symmetric tensor on-shell,

$$j'^\mu(\varphi) - j^\mu(\varphi) \approx \partial_\nu \Sigma^{\nu\mu}(\varphi), \quad \Sigma^{\nu\mu}(\varphi) = -\Sigma^{\mu\nu}(\varphi). \tag{17}$$

Equivalent conserved currents define one and the same integral of motion (16). As for characteristics, the corresponding condition reads

$$N'(\varphi) - N(\varphi) \approx U^\alpha(\varphi)R_\alpha. \tag{18}$$

Equivalent characteristics determine equivalent conserved currents. This establishes a one-to-one correspondence between the equivalence classes of characteristics and conservation laws.

In linear theories, the quadratic conserved currents are the most relevant. Once the conserved current is bilinear in fields, the characteristic is linear. It can be written in the form

$$N(\varphi) = N(\partial)\varphi, \tag{19}$$

where *N* is a matrix differential operator, which we call the operator of characteristics. The defining condition for the operator of characteristic reads

$$N^\dagger(\partial)M(\partial) + M^\dagger(\partial)N(\partial) = 0. \tag{20}$$

The formula can be applied to both variational and non-variational theories. Trivial operators of characteristics read

$$N_{\text{triv}}(\partial) = G(\partial)M(\partial) + \sum_A (Z^\dagger)_A(\partial)U^A(\partial), \quad G^\dagger = -G. \tag{21}$$

In the class of variational models' conditions (20) and (21) take the form of (13) and (14). This brings us to the identification of variational symmetries and characteristics,

$$L(\partial) \equiv N(\partial). \tag{22}$$

The relationship between the variational symmetries and conserved currents is obtained from (20) and (21) using the integration by parts by means of equation (5),

$$j = j_M(\varphi, N\varphi), \tag{23}$$

(the on-shell vanishing terms are omitted). In fact, (22) and (23) represent one of the possible formulations of the classical Noether theorem [1].

As we can see from the above, the classical Noether theorem essentially uses two distinct facts: the relationship between characteristics and conserved currents (23), and the connection between symmetries and characteristics (22). The first of these facts holds true for the variational and non-variational models, while the second uses the presence of action functional. We introduce formula (22) because it allows us to extend the Noether theorem beyond the scope of variational dynamics. A crucial ingredient of generalization of the Noether theorem is the Lagrange anchor.

## 3. Lagrange Anchor and Generalization of the Noether Theorem

The matrix differential operator *V* is called a Lagrange anchor[3] of the linear theory (2) if the following relation is satisfied:

$$V^\dagger(\partial)M^\dagger(\partial) - M(\partial)V(\partial) = 0. \tag{24}$$

The defining equation for the Lagrange anchor has a lot of trivial solutions of the form

$$V_{\text{triv}}(\partial) = G(\partial)M(\partial) + \sum_\alpha R_\alpha(\partial)U^\alpha(\partial), \quad G^\dagger = G, \tag{25}$$

where *G* is a self-adjoint operator, and $U^\alpha$ are arbitrary operators. The trivial Lagrange anchors do not contain valuable information about the dynamics of the theory, and they are also useless for the connection of symmetries and conserved quantities. We systematically ignore them.

The Lagrange anchor is called transitive if it has zero kernel in the class of matrix differential operators,

$$V(\partial)K(\partial) = 0 \quad \Leftrightarrow \quad K(\partial) = 0. \tag{26}$$

We mostly consider transitive Lagrange anchors. The variational models admit the canonical Lagrange anchor, *V=id*, which is transitive. In non-variational theories, some examples of transitive Lagrange anchors can be found in [32,33].

The Lagrange anchor connects symmetries and conserved quantities. If *N* is the characteristic of a conserved quantity, and *V* is a Lagrange anchor, the corresponding symmetry reads

$$L(\partial) = V(\partial)N(\partial). \tag{27}$$

conditions (20) and (24) ensure that the right-hand side of this expression is symmetry in the sense of condition (10). Once the variational theory is equipped with the canonical Lagrange anchor, the identity map connects characteristics and symmetries,

$$L(\partial) = N(\partial), \quad V = id. \tag{28}$$

This is the standard Noether correspondence between symmetries and conserved quantities (22), (23). If *V* is non-canonical, the symmetries and conserved quantities are connected in a non-canonical way.

Relation (27) can be lifted at the level of classes of equivalence of non-trivial symmetries and conserved quantities. To ensure this fact, we should prove that the trivial characteristics are mapped into trivial symmetries. The following relations are used:

$$\begin{aligned} V(\partial)N_{\text{triv}}(\partial) &= V(\partial)G(\partial)M(\partial) + \sum_\alpha V(\partial)(Z^\dagger)_A(\partial)U^A(\partial) = \\ &= (V(\partial)G(\partial))M(\partial) + \sum_\alpha R_\alpha(\partial)U^\alpha(\partial) = L_{\text{triv}}(\partial). \end{aligned} \tag{29}$$

---

[3] For the general definition of Lagrange anchor see [22].

For the sake of simplicity, we assume that the set of gauge generators is complete. In this case, the identity $M(\partial)V(\partial)Z_A(\partial)U^A(\partial) = 0$ implies $V(\partial)Z_A(\partial)U^A(\partial) = R_\alpha(\partial)U^\alpha(\partial)$ for some $U^\alpha$, summation over repeated indices is implied.

The classes of equivalence of characteristics and conserved currents are in one-to-one correspondence. This means that formula (27) connects the classes of equivalence of conservation laws and symmetries, like the Noether theorem. We call relation (27) the generalization of the Noether theorem. As the requirement of the existence of the Lagrange anchor is less restrictive than the presence of the least-action principle, the generalization of the Noether theorem can be applied to connect symmetries and conserved quantities in non-variational theories.

The matrix differential operator $L$ is called a proper symmetry with respect to the Lagrange anchor $V$ if $L(\partial) = V(\partial)P(\partial)$ for some $P$, and

$$V(\partial)(P^\dagger(\partial)M(\partial) + M^\dagger(\partial)P(\partial)) = 0. \tag{30}$$

the proper symmetries are symmetries in the usual sense. Indeed, by relations (24) and (27), we get

$$M(\partial)V(\partial)P(\partial) = (V^\dagger(\partial)M^\dagger(\partial) - M(\partial)V(\partial))P(\partial) + M(\partial)V(\partial)P(\partial) =$$
$$= V^\dagger(\partial)(M^\dagger(\partial)P(\partial) + P^\dagger(\partial)M(\partial)) - V^\dagger(\partial)P^\dagger(\partial)M = V^\dagger(\partial)P^\dagger(\partial)M(\partial). \tag{31}$$

In the class of variational models equipped with the canonical Lagrange anchor, the proper symmetries are just variational ones. If the Lagrange anchor is transitive, the proper symmetries can be connected with integrals of motion. Applying (26) to (30), we get that $P$ is the characteristic,

$$P^\dagger(\partial)M(\partial) + M^\dagger(\partial)P(\partial) = 0. \tag{32}$$

The characteristic defines the conserved current by formula (16). This establishes a correspondence between proper symmetries and conserved quantities. In the class of variational models equipped with the canonical Lagrange anchor, relation (28) establishes the Noether correspondence between the variational symmetries and integrals of motion. In the class of non-variational theories, the proper symmetries constitute a special subset in the space of all symmetries, which can be connected with conserved quantities.

The connection between symmetries and conserved currents is not necessarily unique for a given system of equations, because multiple Lagrange anchors can be admissible by the model. If this takes place, one and the same conserved current can be associated with several different symmetries. Alternatively, one and the same symmetry can come from different pairs of Lagrange anchors and conserved currents. In the rest of the paper we mostly deal with the variational theories of derived type, which are known to admit multiple Lagrange anchors. In particular, we show that the isometries of space-time can be connected with the series of conserved tensors in this class of models.

**4. Higher-Derivative Theories of Derived Type**

Given a set of fields $\varphi^i$, we introduce the variational primary model without higher-derivatives

$$W_{ij}(\partial)\varphi^j = 0, \quad W^\dagger = W. \tag{33}$$

The model (2) falls into the class of theories of derived type if its wave operator is a constant coefficient polynomial in $W$,

$$M(\alpha;W) = \sum_{p=0}^{n} \alpha_p W^p, \quad \alpha_n \neq 0. \tag{34}$$

The constants $\alpha_0,\ldots,\alpha_n$ distinguish different representatives in the class. The equations of motion (33) come from the least-action principle for the functional

$$S[\varphi(x)] = \frac{1}{2}\int \langle \varphi, W\varphi \rangle d^d x. \tag{35}$$

Each derived theory is defined by a primary operator *W*, and a characteristic polynomial in the complex variable z,

$$M(\alpha;z) = \sum_{p=0}^{n} \alpha_p z^p. \tag{36}$$

The symmetries, characteristics, and Lagrange anchors in the class of derived systems can be systematically obtained from the corresponding quantities of the primary model. Even though the derived quantities do not cover all symmetries, characteristics, and Lagrange anchors, they contain valuable information about dynamics. In the present paper, we use derived conserved quantities to study the stability of higher-derivative models.

Let us explain the details of the construction. Suppose that the primary theory admits the series of variational symmetries with the operator $L(\xi)$,

$$L(\xi) = \sum_{a=1}^{s} \xi^a L_a, \quad [L_a, W] = 0, \quad L_a = -(L_a)^\dagger, \tag{37}$$

where the indices *a* = 1,…,*s* label the generators of symmetry series, and $\xi$'s are the transformation parameters, being real constants. In this case, the Noether theorem (22), (23) associates the conserved currents $j_a$, *a* = 1,…,*s* with these symmetries. The conserved currents are linearly independent if the symmetry generators are linearly independent (modulo trivial symmetries). In what follows in this section, we assume the linear independence of the generators $L_a$ (37). The operators $L_a$ (37) can form a Lie algebra, at least in some models. The Poincare symmetry in the relativistic theories is one of the relevant examples of such kind. We do not specify the structure of the algebra of symmetries because we are mostly interested in the correspondence between the linear spaces of symmetries and conserved quantities.

A single variational symmetry (37) defines the *n*-parameter series of variational symmetries of the action functional (4), (34),

$$N(\beta,\xi;W) = L(\xi)N(\beta;W), \quad N(\beta;W) \equiv \sum_{p=0}^{n-1} \beta_p W^p, \tag{38}$$

where $\beta_p$, *p* = 0,…,*n*-1, are constant parameters. For *p* = 0, this series includes the symmetries (37) of the primary model (33). The other entries in (38) are higher-derivative operators whose origin is a consequence of the derived structure (34) of the equations of motion (2). The symmetries in the series (38) are usually independent, even though it is not a theorem. The general argument is that if all the powers of the primary operator are independent initial data and $L_a$'s are transitive operators, the characteristic (38) cannot be zero on-shell. The exceptions are possible in the class of gauge theories, where some entries of the series (38) can trivialize on the mass-shell with account of gauge symmetries.

The Noether theorem (22), (23) associates the following set of conserved currents with the variational symmetry (38):

$$j(\beta,\xi) = \sum_{p=0}^{n-1} \beta_p j^p(\xi),$$

$$j^p(\xi) = \sum_{q=0}^{n} \alpha_q \left( \sum_{s=1}^{q-p} j_W(L(\xi)W^{p+s-1}\varphi, W^{q-s}\varphi) + \sum_{s=1}^{p-q} j_W(L(\xi)W^{p-s}\varphi, W^{q+s-1}\varphi) \right). \tag{39}$$

The derivation of this formula uses integration by parts by means of equation (5). All the on-shell vanishing contributions are omitted in (39). In this set, $j^0$'s represents the canonical conserved quantities of the derived model (4), (34) that are connected with the original symmetry (37) of the primary model (33). The other conserved currents, $j^p$, *p* = 1,…,*n*-1, come from the higher-symmetries in the set (38). The conditions of linear independence of conserved currents are the same as that of symmetries.

The Lagrange anchors in the class of derived theories are just polynomials in the primary operator,

$$V(\gamma;W) = \sum_{p=0}^{n-1} \gamma_p W^p. \tag{40}$$

The higher-powers of $W$ do not define new non-trivial Lagrange anchors, see equation (25). The canonical Lagrange anchor is included in this series for

$$\beta_0 = 1, \quad \beta_1 = \beta_2 = ... = \beta_{n-1} = 0. \tag{41}$$

The Lagrange anchor (40) connects characteristic (38) with symmetry of the derived theory (34) by the rule (27). The resulting symmetry is the original one if

$$V(\gamma;W)N(\beta,\xi;W) = L(\xi) + \text{trivial symmetry}. \tag{42}$$

We are interested in the problem of connecting of conserved quantities with the original symmetry because it is relevant for studying stability.

Let us first consider the theories without gauge symmetries. In this case, the relevant trivial symmetry in the right-hand side of (42) has the form

$$L_{\text{triv}} = K(\delta;W)M(\alpha;W)L(\xi), \quad K(\delta;W) = \sum_{p=0}^{n-2} \delta_p W^p, \tag{43}$$

with $K$ being some polynomial in $W$. To meet (42), it is sufficient to impose the condition

$$V(\gamma;W)N(\beta;W) = id + K(\delta;W)M(\alpha;W). \tag{44}$$

If all the powers of the characteristic operator are independent, the formula is equivalent to the relation between the characteristic polynomials of the involved quantities

$$V(\gamma;z)N(\beta;z) - K(\delta;z)M(\alpha;z) = 1. \tag{45}$$

The Bezout Lemma states that the problem has a unique solution for $V$ and $K$ if, and only if, $N$ and $M$ have no common roots. Once two general polynomials have no common roots, almost all the conserved quantities in the series are related to the original symmetry.

In the class of gauge models, some additional symmetry can trivialize on-shell. The relevant trivial symmetry in the right-hand side of (42) can be chosen in the form

$$L_{\text{triv}} = \left( K(\delta;W)M(\alpha;W) + \sum_{k=1}^{s} K_s(W)M_s(W) \right) L(\xi), \tag{46}$$

where $M_s$ are additional generators of trivial symmetries. In principle, the wave operator and $M_s$ can be dependent in some models. We should ignore $M$ in the formulas below in these circumstances. The analogue of relations (44), (45) reads

$$V(\gamma;W)N(\beta;W) - K(\delta;W)M(\alpha;W) - \sum_{k=1}^{s} K_s(W)M_s(W) = id; \tag{47}$$

$$V(\gamma;z)N(\beta;z) - K(\delta;z)M(\alpha;z) - \sum_{k=1}^{s} K_s(z)M_s(z) = 1. \tag{48}$$

These equations are consistent if $N$ and at least one characteristic polynomial of the trivial gauge symmetry generators $M$, $M_s$, have no common roots. This condition is less restrictive than in the case of non-gauge models.

**5. Stability of Higher-derivative Dynamics**

In this section, we address the problem of establishing a relationship between the bounded-conserved quantities and the time translation symmetry. This problem needs more accurate investigation because bounded-conserved quantities are not necessarily general representatives of conserved current series, while general conserved currents may define unbounded-conserved charges. We proceed in two steps. First, the bounded-conserved quantities in the series are identified. After that, we address the issue of connecting bounded-conserved quantities with the space-time translation. Throughout the section we assume that $L(\xi) = \xi^\mu \partial_\mu$.

Let us discuss the structure of conserved quantities in the series (39). We introduce the factorization of the wave operator (34) in the form

$$M(\alpha;W) = \alpha_n \prod_{k=1}^{r} (W - \lambda_k)^{m_k}, \tag{49}$$

where the numbers $\lambda_k$ and $m_k$ label different roots and their multiplicities. The integer $r$ is the total number of different roots. We assume that all the quantities $\lambda_p$ are real numbers. We do not consider complex roots because the corresponding derived theory is always unstable.

Using factorization (49), we define the new dynamical variables absorbing the derivatives of the original field by the receipt of Pais and Uhlenbeck [26]:

$$\varphi_k = \Lambda_k(\partial)\varphi, \quad \Lambda_k(\partial) = \alpha_n \prod_{i=1, i \neq k}^{r} (W - \lambda_i)^{m_i}, \quad k = 1,...,r. \tag{50}$$

We term the new quantities as components. By construction, the original field $\varphi$ can be expressed on-shell as the linear combination:

$$\varphi = \sum_{k=1}^{r} C_k(\zeta;W)\varphi_k, \quad C_k(\zeta;W) = \sum_{k=0}^{m_k - 1} \zeta_k W^k. \tag{51}$$

The following system of equations is valid for the components:

$$(W - \lambda_k)^{m_k} \varphi_k = 0, \quad k = 1,...,r. \tag{52}$$

It ensures that the components, which are associated with different roots of characteristic equation, are independent degrees of freedom. The one-to-one correspondence between the solutions of systems (2) and (52) is established by formulas (50) and (51).

The system of equations for components comes from the action functional

$$S = \sum_{k=1}^{r} S_k[\varphi_k(x)], \quad S_k[\varphi_k(x)] = \frac{1}{2} \int \langle \varphi_k, (W - \lambda_k)^{m_k} \varphi_k \rangle d^d x. \tag{53}$$

Once $L(\xi)$ is a set of symmetries of the primary model, the following transformations are variational symmetries:

$$\delta_\xi \varphi_k = (W - \lambda_k)^l L(\xi) \varphi_k, \quad k = 1,...,r, \quad l = 0,...,m_k - 1, \tag{54}$$

with no sum in $k$. There are $n$ independent symmetries in this set. In terms of the original field set, these transformations correspond to the derived symmetries (38). The following conserved currents are associated with them:

$$j^{k;m_k - l - 1}(\xi) = \sum_{s=1}^{m_k - l} j_{W - \lambda_k} (L(\xi)(W - \lambda_k)^{l+s-1} \varphi_k, (W - \lambda_k)^{m_k - s} \varphi_k),$$

$$k = 1,...,r, \quad l = 0,...,m_k - 1. \tag{55}$$

In this set, the leading representatives $j^{k;m_k-l}$ are the canonical energy-momentum currents of the components, while the others are independent quantities. It is obvious that (55) are linear combinations of conserved tensors (39).

Once $\xi$ is an arbitrary constant vector, the conserved quantities are tensors. The second-rank conserved tensors $(T^{k;l})^{\mu\nu}$ are defined by the rule

$$(j^{k;l}(\xi))^\mu = \xi_\nu (T^{k;l})^{\mu\nu}. \tag{56}$$

As for the structure of conserved tensors (56), the following observation is relevant. The leading representatives in the conserved tensor series,

$$(T^{k;0})^{00} = (j_{W-\lambda_k}(\partial^0 (W-\lambda_k)^{m_k-1}\varphi_k, (W-\lambda_k)^{m_k-1}\varphi_k))^0, \tag{57}$$

have bounded from below 00-component if the canonical energy of the primary model is bounded on-shell

$$(T^{k;0})^{00} \geq 0 \quad \Leftrightarrow \quad T^{00}_{W-\lambda_k}(\varphi) \equiv (j_{W-\lambda_k}(\partial^0\varphi,\varphi))^0 \geq 0, \quad \forall \varphi. \tag{58}$$

These conditions can be satisfied in many cases. For example, it is sufficient to assume

$$T^{00}_W(\varphi) \geq 0, \quad T^{00}_{\lambda_k}(\varphi) = -\lambda_k \langle \varphi, \varphi \rangle \geq 0. \tag{59}$$

The other contributions in (55) are not bounded from below because they are linear in the variable

$$(W-\lambda_k)^{m_k-1}\varphi_k. \tag{60}$$

Once this quantity is an initial data of the model, the conserved tensors (56) cannot have bounded 00-component for $l > 0$.

The observations above can be summarized as follows. If the primary theory is stable, the leading terms in the conserved tensor set (56) have bounded 00-components, while the other contributions are unbounded unless they vanish due to gauge symmetries or constraints. The bounded series of conserved currents in unconstrained theory has the form

$$(j^+)^\mu(\beta,\xi) = \sum_{k=1}^r \beta_k \xi_\nu (T^{k;0})^{\mu\nu}. \tag{61}$$

In the case of multiple roots of the characteristic equation, this subseries is special because it involves only some initial data (60). Such conserved quantities cannot ensure the classical stability of the model. This brings us to the conclusion that non-singular theories of derived type with multiple roots should be unstable. This observation is supported by the models of the Pais–Uhlenbeck oscillator and the higher-derivative scalar field, where the models with resonance are unstable.

Let us now show that theories with multiple roots cannot be stable at the quantum level. This amounts to the fact that the bounded-conserved quantity cannot be connected with the time translation symmetry. The characteristic for the bounded-conserved tensor reads

$$N^+(\beta,\xi;W) = \sum_{k=1}^r L(\xi)\beta_k (W-\lambda_k)^{m_k-1}\Lambda_k(\lambda;W), \tag{62}$$

where $\Lambda$'s were introduced in (50). Given the Lagrange anchor (40), the corresponding symmetry is (42). This expression defines the space-time translation if

$$L(\xi)V(\gamma;W)N^+(\beta;W) = L(\xi) + \text{trvial symmetry}. \tag{63}$$

Assuming that no gauge symmetries are present in the model, and $L(\xi)$ is a transitive operator (26), we conclude that

$$V(\gamma;W)N^+(\beta;W) - K(\delta;W)M(\alpha;W) = id, \tag{64}$$

where $K$ is some polynomial. Once all the powers of $W$ are linearly independent, this equation is equivalent to the following relation between characteristic polynomials of conserved quantities:

$$V(\gamma;z)N^+(\beta;z) - K(\delta;z)M(\alpha;z) = 1. \tag{65}$$

By the Bezout Lemma, this condition is consistent for fixed $N^+$ and $M$ if, and only if, these polynomials have no common roots. On the other hand, each multiple root is common for $M$ and $N^+$. Hence, the bounded-conserved quantity can only be connected with space-time translations if all the roots of the characteristic equation are simple. The Pais–Uhlenbeck oscillator and higher-derivative scalar field models again serve as demonstrations for this observation.

The case of gauge theories needs more accurate consideration. There are two important points: unbounded contributions in (55) can trivialize on the mass-shell, and the relation between symmetries and conserved quantities is more relaxed. This allows us to connect a bounded-conserved tensor with the time translation symmetry even if the characteristic polynomial has multiple roots. We illustrate this possibility in the extended Chern–Simons theory in Subsection 6.3.

## 6. Examples

*6.1. Fourth-order Pais–Uhlenbeck Oscillator.*

The Pais–Uhlenbeck oscillator of the fourth-order is a theory of a single dynamical variable $x(t)$ with the action functional

$$S[x(t)] = \frac{1}{2}\int \left( x\left(\frac{d^2}{dt^2} + \omega_1^2\right)\left(\frac{d^2}{dt^2} + \omega_2^2\right)x \right)dt, \quad (66)$$

where the frequencies $\omega_1$, $\omega_2$ are parameters of the model. The Euler–Lagrange equation for the action functional is of derived type,

$$\frac{\delta S}{\delta x} = \left(\frac{d^2}{dt^2} + \omega_1^2\right)\left(\frac{d^2}{dt^2} + \omega_2^2\right)x = 0, \quad (67)$$

with the primary operator being the second time derivative. The squares of frequencies determine the roots of the characteristic polynomial (36), which can be equal or different. The order of the characteristic polynomial equals two.

If the frequencies of the Pais–Uhlenbeck oscillator are different, two integrals of motion (56) are admissible by the model,

$$J^k = \frac{1}{2(\omega_1^2 + \omega_2^2)}\left((\ddot{x} + \omega_k^2 \dot{x})^2 + (\omega_1^2 + \omega_2^2 - \omega_k^2)(\ddot{x} + \omega_k^2 x)^2\right), \quad k = 1, 2. \quad (68)$$

These conserved quantities are obviously bounded from below. The bounded subseries of conserved quantities (61) is the linear combination of these expressions,

$$J^+ = \sum_{k=1}^{2} \beta_k J^k, \quad \beta_k > 0. \quad (69)$$

The characteristic for the bounded-conserved quantity reads

$$N^+\left(\beta, \frac{d^2}{dt^2}\right) = \sum_{k=1}^{2} \frac{\beta_k}{\omega_1^2 + \omega_2^2}\left(\frac{d^2}{dt^2} + \omega_k^2\right)\frac{d}{dt}. \quad (70)$$

The Lagrange anchor,

$$V\left(\frac{d^2}{dt^2}\right) = \frac{(\omega_1^2 + \omega_2^2)}{(\omega_2^2 - \omega_1^2)^2}\left(\frac{\beta_1 + \beta_2}{\beta_1\beta_2}\frac{d^2}{dt^2} + \frac{\beta_1\omega_2^2 + \beta_2\omega_1^2}{\beta_1\beta_2}\right), \quad (71)$$

connects the conserved quantity (69) with the time translation symmetry whenever the product $\beta_1\beta_2$ is nonzero. This gives an alternative proof of the stability of the Pais–Uhlenbeck oscillator with different frequencies.

In the case of resonance $\omega_1 = \omega_2 = \omega$, the Pais–Uhlenbeck oscillator has two conserved quantities, only one of which is bounded from below,

$$J^1 = \frac{1}{2\omega^2}\left((\ddot{x} + \omega^2 \dot{x})^2 + (\ddot{x} + \omega^2 x)^2\right), \quad J^2 = \frac{1}{2\omega^2}(2\dot{x}\,\dddot{x} - \ddot{x}^2) + x^2 + \frac{1}{2}\omega^2 x^2. \quad (72)$$

The series of bounded-conserved quantities (61) includes a single entry $J^1$. The characteristic (62) for this conserved quantity reads

$$N^+\left(\frac{d^2}{dt^2}\right) = \frac{1}{\omega^2}\left(\frac{d^2}{dt^2} + \omega^2\right)\frac{d}{dt}. \quad (73)$$

It has the common root $\omega^2$ with the characteristic polynomial of the wave operator (67). This means that the bounded integral of motion cannot be connected with the time translation symmetry in the model with resonance.

The results of this subsection show that the fourth-order Pais–Uhlenbeck oscillator is stable if its frequencies are different. This result confirms the general observation about the connection of structure of the roots of characteristic polynomial and its stability.

*6.2. Higher-derivative Scalar Field.*

Consider the theory of real scalar field $\varphi(x)$ on $d$-dimensional Minkowski space with the action functional

$$S[\varphi(x)] = \frac{1}{2}\int \varphi\left(\prod_{p=1}^{n}\left(\frac{\partial^2}{\partial x^\mu \partial x_\mu} + \mu_p^2\right)\right)\varphi d^d x. \tag{74}$$

In this formula, the real numbers $\mu_p$ determine the spectrum of masses in the theory. The equations of motion belong to derived type, with the primary operator being d'Alembertian,

$$\frac{\delta S}{\delta \varphi} = \prod_{p=1}^{n}\left(\frac{\partial^2}{\partial x^\mu \partial x_\mu} + \mu_p^2\right)\varphi = 0, \quad W = \frac{\partial^2}{\partial x^\mu \partial x_\mu}. \tag{75}$$

The primary model is the theory of free mass-less scalar field,

$$\frac{\partial^2}{\partial x^\mu \partial x_\mu}\varphi = 0. \tag{76}$$

This theory is invariant under the Poincare symmetries, including the space-time translations.

The structure of conserved quantities in the model depends on the values of the roots of the characteristic polynomial. Once all roots are different, all the dynamical degrees of freedom are scalars with different masses. The lower order formulation (52) for the model reads

$$\left(\frac{\partial^2}{\partial x^\mu \partial x_\mu} + \mu_p^2\right)\varphi_p = 0, \quad \varphi_p = \prod_{q=1, q\neq p}^{n}\left(\frac{\partial^2}{\partial x^\mu \partial x_\mu} + \mu_q^2\right)\varphi, \quad p=1,\ldots,n. \tag{77}$$

The conserved tensors (56) are just energies of the components,

$$(T^p)^{\mu\nu} = \partial^\mu \varphi_p \partial^\nu \varphi_p - \frac{1}{2}\eta^{\mu\nu}(\partial^\rho \varphi_p \partial_\rho \varphi_p - \mu_p^2 \varphi_p^2), \quad p=1,\ldots,n. \tag{78}$$

It is clear that all of these quantities have bounded from below 00-component. The subseries of bounded-conserved quantities (61) have the form

$$(j^+)^\mu(\beta;\xi) = \sum_{p=1}^{n}\beta_p \xi_\nu (T^p)^{\mu\nu}, \quad \beta_p > 0. \tag{79}$$

By the general theorem above, all of these quantities can be connected with the space-time translations by the appropriate Lagrange anchor. We derive the explicit expression for such a Lagrange anchor in Appendix A.[4]

If multiple roots are admissible for the characteristic polynomial, two or more conserved quantities are related with one and the same root. Below, we give expressions for additional integrals of motion in the simplest option, where only one root has multiplicity two, and all the other roots are simple. Without loss of generality we assume that

$$\mu_{n-1} = \mu_n = \mu. \tag{80}$$

In this case, the system (52) reads

$$\left(\frac{\partial^2}{\partial x^\mu \partial x_\mu} + \mu_p^2\right)\varphi_p = 0, \quad \varphi_p = \prod_{q=1, q\neq p}^{n}\left(\frac{\partial^2}{\partial x^\mu \partial x_\mu} + \mu_q^2\right)\varphi, \quad p=1,\ldots,n-2, \tag{81}$$

---

[4] For the fourth-order theory (case $n=2$) such a Lagrange anchor has been first introduced in [21].

$$\left(\frac{\partial^2}{\partial x^\mu \partial x_\mu} + \mu^2\right)^2 \varphi_{n-1} = 0, \quad \varphi_{n-1} = \prod_{q=1}^{n-2}\left(\frac{\partial^2}{\partial x^\mu \partial x_\mu} + \mu_q^2\right)\varphi. \tag{82}$$

The components $\varphi_p$, $p = 1,\ldots,n$-2 are usual scalar fields, while $\varphi_{n-1}$ obeys higher-derivative equations. For simple roots the conserved tensors have the form (78), we do not repeat the expressions for them. Two conserved tensors are associated with the multiple root,

$$(T^{n-1;0})^{\mu\nu} = \frac{1}{\mu^2}\left(\partial^\mu \tilde\varphi_{n-1} \partial^\nu \tilde\varphi_{n-1} - \frac{1}{2}\eta^{\mu\nu}(\partial^\rho \tilde\varphi_{n-1}\partial_\rho \tilde\varphi_{n-1} - \mu^2 \tilde\varphi_{n-1}^2)\right), \quad \tilde\varphi_{n-1} \equiv (\partial_\mu \partial^\mu + \mu^2)\varphi_{n-1};$$

$$(T^{n-1;1})^{\mu\nu} = \frac{1}{\mu^2}\Big(\partial^\mu \varphi_{n-1}\partial^\nu \partial_\rho \partial^\rho \varphi_{n-1} - \partial^\mu \partial_\rho \varphi_{n-1}\partial^\nu \partial^\rho \varphi_{n-1} + 2\mu^2 \partial^\mu \varphi_{n-1}\partial^\nu \varphi_{n-1} - \tag{83}$$

$$-\frac{1}{2}\eta^{\mu\nu}(-\partial^\rho \partial^\tau \varphi_{n-1}\partial_\rho \partial_\tau \varphi_{n-1} + \mu^2 \partial^\rho \varphi_{n-1}\partial_\rho \varphi_{n-1} - \mu^4 \varphi_{n-1}^2)\Big).$$

The bounded-conserved quantity reads

$$(j^+)^\mu(\beta;\xi) = \sum_{p=1}^{n-1}\beta_p \xi_\nu (T^{p;0})^{\mu\nu}, \quad \beta_p > 0. \tag{84}$$

The characteristic for the bounded-conserved tensor has the form

$$N^+\left(\beta,\xi;\frac{\partial^2}{\partial x^\mu \partial x_\mu}\right) = \left(\frac{\partial^2}{\partial x^\mu \partial x_\mu} + \mu^2\right)\left(\sum_{p=1}^{n-1}\left(\beta_p \xi_\mu \partial^\mu \prod_{q=1,q\neq p}^{n-1}\left(\frac{\partial^2}{\partial x^\mu \partial x_\mu} + \mu_q^2\right)\right)\right). \tag{85}$$

Its characteristic polynomial,

$$N^+(\beta;z) = (z+\mu^2)\sum_{p=1}^{n-1}\left(\beta_p \prod_{q=1,q\neq p}^{n-1}(z+\mu_q^2)\right), \tag{86}$$

has the common root $-\mu^2$ with the characteristic polynomial of the theory. In doing so, the bounded-conserved quantity cannot be connected with the time translations. This demonstrates that the free higher-derivative scalar field theory with different masses is stable, while the multiple roots indicate the instability of the model.

*6.3. Extended Chern–Simons Model.*

Consider the theory of the vector field $A = A_\mu(x)dx^\mu$ on 3*d* Minkowski space with the action functional

$$S[A(x)] = \frac{1}{2}\int\left(A,\sum_{p=1}^{n}\alpha_p(*d)^p A\right)d^3x, \tag{87}$$

where the round brackets denote the standard inner product of differential forms, * is the Hodge dual, and *d* is the de-Rham differential. The parameters of the model are the constants $\alpha_1,\ldots,\alpha_n$. The action of the Chern–Simons operator on the vector field is determined by the relation

$$(*dA)_\mu dx^\mu = \varepsilon_{\mu\nu\rho}dx^\mu \partial^\nu A^\rho. \tag{88}$$

Here, $\varepsilon$ is the 3*d* Levi-Civita symbol, with $\varepsilon_{012} = 1$. The Euler–Lagrange equations for the model have the form

$$\frac{\delta S}{\delta A} = \sum_{p=1}^{n}\alpha_p(*d)^p A = 0. \tag{89}$$

The primary operator of the theory is the Chern–Simons one, see equation (88). The primary theory for the model is the usual abelian Chern–Simons theory.

The primary operator (88) is Poincare-invariant, so that the space-time translations are symmetries of the model. The series of derived symmetries reads

$$N(\beta,\xi;*d) = \sum_{p=0}^{n-1} \xi^\mu \beta_p (*d)^p \partial_\mu. \tag{90}$$

The corresponding set of conserved tensors has the form

$$T^{\mu\nu}(\beta) = \frac{1}{2} \sum_{p,q=0}^{n-1} C_{p,q}(\alpha,\beta)(F^{(p)\mu}F^{(q)\nu} + F^{(p)\mu}F^{(q)\nu} - \eta^{\mu\nu}\eta_{\rho\sigma}F^{(p)\rho}F^{(q)\sigma}), \tag{91}$$

all of the space-time indices are raised and lowered by the Minkowski metric, and the following notation is used:

$$F^{(p)} = (*d)^p A, \quad p = 0,...,n-1. \tag{92}$$

The quantity $C(\alpha,\beta)$ (91) is the Bezout matrix of the characteristic polynomial of the model and the characteristic polynomial of the symmetry. It is defined by the generating relation,

$$C_{p,q}(\alpha,\beta) = \frac{\partial^{p+q}}{\partial^p z \partial^q u}\left(\frac{M(z)N(u) - M(u)N(z)}{z - u}\right)\bigg|_{z=u=0},$$

$$M(z) \equiv \sum_{p=1}^{n} \alpha_p z^p, \quad N(z) \equiv \sum_{p=0}^{n-1} \beta_p z^{p+1}. \tag{93}$$

where $z$ and $u$ are two independent variables. The individual conserved tensors in the set (91) can be found by the following receipt:

$$(T^p)^{\mu\nu} = \frac{\partial T^{\mu\nu}(\beta)}{\partial \beta_p}, \quad p = 0,...,n-1. \tag{94}$$

In this set, the quantities $(T^p)^{\mu\nu}$, $p = 0,...,n-2$, are independent, while

$$(T^{n-1})^{\mu\nu} = -\sum_{p=0}^{n-2} \frac{\alpha_p}{\alpha_n}(T^p)^{\mu\nu}. \tag{95}$$

We mention this fact to illustrate possible dependence among conserved quantities in the class of gauge models. We also notice that the set of the conserved currents of the extended Chern–Simons model (87) was introduced in the work [25], while the compact form (91–94) is proposed in [34].

The structure of the conserved currents can be studied along the lines of the previous section. The components (50) are introduced by the standard rule,

$$A_k = \Lambda_k(\partial)A, \quad \Lambda_k(\partial) = \alpha_n \prod_{i=1, i \neq k}^{r} (*d - \lambda_i)^{m_i}, \quad k = 1,...,r. \tag{96}$$

The corresponding conserved quantities (55) are determined by formulas (91–94) with

$$M = (z - \lambda_k)^{m_k}, \quad Q = (z - \lambda_k)^l, \quad k = 1,...,r, \quad l = 0,...,m_k - 1. \tag{97}$$

These conserved currents can be found for each particular value of the roots. Without loss of generality we assume that the zero root corresponds to $k = 1$, while all the other numbers $\lambda_k$ are non-zero.

As for the structure of the conserved current, we mention that the contribution with the highest derivative has the form

$$(T^{k;l})^{\mu\nu} = \frac{1}{2}(-\lambda)^{m_k}(F_k^{(m_k-1)\mu}F_k^{(l)\nu} + F_k^{(m_k-1)\nu}F_k^{(l)\mu} - \eta^{\mu\nu}\eta_{\rho\sigma}F_k^{(m_k-1)\rho}F_k^{(l)\sigma}) + ...,$$

$$F_k^{(l)} = (*d - \lambda_k)^l \Lambda_k A, \quad k = 1,...,r, \quad l = 0,...,m_k - 1. \tag{98}$$

The dots denote all the contributions with lower derivatives. These conserved quantities cannot be bounded unless $l = m_k-1$. For $l = m_k-1$, expressions (98) are exact with no dotted terms, and the corresponding conserved quantities are bounded.

As is seen from formulas (98), the leading representatives $T^{k;0}$ in the conserved-quantity series are bounded. For simple zero roots the corresponding conserved quantity is the Chern–Simons energy, which is trivial. The additional conserved currents are associated with multiple roots. In the case of non-zero root, all these conserved quantities are independent and unbounded. In the case of zero roots, the additional quantity has a bounded 00-component[5], while all the other independent entries are independent and unbounded. This means that all of the additional conserved quantities in the set (56) are unbounded if the multiple zero root has a multiplicity greater than two. The subseries (61) of conserved quantities with the bounded 00-component has the form

$$(j^+)^\mu(\beta;\xi) = \sum_{k=2}^{r}\beta_{k;0}\xi_\nu(T^{k;0})^{\mu\nu} + \beta_{1;1}\xi_\nu(T^{1;1})^{\mu\nu}. \tag{99}$$

The characteristic of the bounded-conserved tensor series reads

$$N^+(\beta;\xi) = -\sum_{k=2}^{r}\text{sgn}(\lambda_k)\beta_{k;0}\Lambda_k(*d)L(\xi) + \beta_{1;1}(*d)\Lambda_1(*d)L(\xi). \tag{100}$$

This is not the general representative of a characteristic series (90), because it involves fewer entries.

Let us discuss the stability of the model. The series (40) of the Lagrange anchors for the extended Chern–Simons model has the following form:

$$V(\gamma;*d) = \sum_{p=0}^{n-1}\gamma_p(*d)^p, \tag{101}$$

with $\gamma_0,\ldots,\gamma_{n-1}$ being real numbers. All the entries of the series are non-trivial. The Lagrange anchor (101) takes the characteristic (100) into a symmetry by the rule (27). This symmetry is the space-time translation if the condition (42) is satisfied. The general trivial symmetry in the considered case reads

$$L_{\text{triv}}(\xi;*d) = L_\xi K(\delta;*d)M^-(\alpha;*d), \quad M^-(\alpha;*d) = \sum_{p=1}^{n}\alpha_p(*d)^{p-1}, \tag{102}$$

where $K$ is an arbitrary polynomial in $*d$. To ensure trivialization of this symmetry, one can use the Cartan formula for the Lie derivative,

$$L_\xi K(\delta;*d)M^-(\delta;*d) = (i_\xi d + di_\xi)K(\delta;*d)M^-(\delta;*d) =$$
$$= i_\xi * K(\delta;*d)M(\delta;*d) + di_\xi K(\delta;*d)M^-(\delta;*d). \tag{103}$$

Substituting (100), (101), and (102) into (47), we get the equation

$$V(\gamma;*d)N^+(\beta;*d) - K(\delta;*d)M^-(\alpha;*d) = id, \tag{104}$$

where the constants $\gamma$, $\delta$ are unknown. In terms of characteristic polynomials of the Lagrange anchor and symmetry, the following equation is relevant:

$$V(\gamma;z)N^+(\beta;z) - K(\delta;z)M^-(\alpha;z) = 1. \tag{105}$$

It is consistent if $N^+$ and $M^-$ have no common roots. The following restriction on the multiplicity of roots is implied:

$$m_1 = 1,2, \quad m_k = 0,1, \quad k = 2,\ldots,r. \tag{106}$$

In this case, all the non-zero roots should be simple and real, and the zero root should have a multiplicity of one or two. This requirement is less restrictive than the absence of a common root between $N^+$ and the characteristic polynomial of the derived theory, which has the place in case of non-gauge systems.

## 7. Conclusions

---

[5] It is the canonical energy of *3d* electrodynamics, which is known to be bounded.

In this article, we have studied the stability of the class of relativistic higher-derivative theories of derived type from the viewpoint of a more general correspondence between symmetries and conserved quantities, which is established by the Lagrange anchor. Assuming that the wave operator of the linear model is the *n*-th-order polynomial in the lower-order operator, we have obtained the following results. First, we observed that *n*-parameter series of second-rank conserved tensors and Lagrange anchors are admissible by the derived model. The canonical energy, which is unbounded, is included into the series in all of the instances. The other integral of motion can be bounded or unbounded depending on the structure of the roots of the characteristic polynomial. The general conserved tensor in the series can be connected with the space-time symmetries by an appropriate Lagrange anchor. Second, we studied the stability of higher-derivative dynamics from the viewpoint of correspondence between the time translation symmetry and bounded-conserved quantities. It has been observed that this relationship can be established if all of the roots of the characteristic polynomial are real and simple. For multiple roots, bounded-conserved quantities are admissible, but they are unrelated with the time translation.

Our stability analysis is applicable to free models. The real issue is the stability of higher-derivative dynamics at the non-linear level. This subject has been studied in many works and we cite recent papers [3–5,35,36] and references therein. The method of proper deformation [37] has been proposed to systematically deform the equations of motion and conserved quantities. Once this method uses the Lagrange anchor construction it can be used to deform bounded-conserved tensors, which are found in this paper. In doing so, the stability of linear higher-derivative models, which are studied in the present paper, can be extended at the non-linear level.

**Funding:** This research was funded by the state task of Ministry of Science and Higher Education of Russian Federation, grant number 3.9594.2017/8.9.

**Acknowledgments:** I thank S.L. Lyakhovich, A.A. Sharapov and V.A. Abakumova for valuable discussions on the subject of this research. I benefited from the valuable comments of two anonymous referees that helped me to improve the initial version of the manuscript. I also thank my family who supported me at all stages of this work.

**Conflicts of Interest:** The funders had no role in the design of the study, in the collection, analyses, or interpretation of data, in the writing of the manuscript, or in the decision to publish the results.

**Appendix A**

In this appendix, we demonstrate that the general bounded-conserved tensor (79) in the theory of the higher-derivative scalar field with different masses (74) can be connected with the space-time translations by the appropriate Lagrange anchor. To solve the problem, we find the characteristic polynomials $V(\gamma,z)$ and $K(\delta,z)$ that satisfy condition (45). In this case, the Lagrange anchor is defined by the formula (40), with $W$ being the d'Alembertian.

At first, we identify all the ingredients in (45), including the characteristic polynomials of the wave operator and characteristic,

$$M(\mu;z) = \prod_{p=1}^{n}(z+\mu_p^2), \quad N^+(\beta;z) = \sum_{p=1}^{n}\beta_p\Lambda_p(z), \quad V(\gamma;z) = \sum_{p=1}^{n}\gamma_p\Lambda_p(z),$$

$$K(\delta;z) = \prod_{p=0}^{n-2}\delta_p z^p, \quad \Lambda_p(z) \equiv \prod_{q=1,q\neq p}^{n}(z+\mu_q^2). \tag{A.1}$$

Here, the expressions $M(\mu;z)$ and $N(\beta;z)$ are fixed by the model parameters and the choice of the selected representative in the conserved tensor series, while $\gamma$, $\delta$ are unknown constants. The quantities $\Lambda_p(z)$ denote characteristic polynomials of operators (50). By construction, $\Lambda_p(z)$ form the basis in the space of polynomials of order $n-1$ in the variable $z$. In this setting, the chosen ansatz for $V(\gamma;z)$ is a different parameterization of Lagrange anchor series (40).

The defining equation (45), (A.1) for the characteristic polynomial $V(\gamma;z)$ has the form

$$\sum_{p=1}^{n}\sum_{q=1}^{n}\beta_p\gamma_q\Lambda_p(z)\Lambda_q(z) - K(\delta;z)M(\mu;z) = 1. \tag{A.2}$$

From here, the polynomial $K(\delta,z)$ is immediately found,

$$K(\delta;z) = \sum_{p=1}^{n}\sum_{q=1}^{n} \beta_p \gamma_q \Lambda_{pq}(z), \quad \Lambda_{pq}(z) = \prod_{r \neq p,q}(z + \mu_r^2). \tag{A.3}$$

As $K(\delta,z)$ is the fraction of $V(\gamma;z) \cdot N(\beta,z)$ and $M(\mu;z)$, no restrictions on the parameters $\gamma$ in the Lagrange anchor appear at this step. After that we estimate left- and right-hand sides of the relation (A.2) for

$$z = -\mu_p^2, \quad p = 1,...,n, \tag{A.4}$$

being the roots of characteristic polynomial. We arrive to the following system of equations:

$$\beta_p \gamma_p \Lambda_p^2(-\mu_p^2) = 1, \quad p = 1,...,n. \tag{A.5}$$

From here, the parameters $\gamma$ are determined,

$$\gamma_p = \frac{1}{\beta_p \Lambda_p^2(-\mu_p^2)}. \tag{A.6}$$

This solution is well-defined since $\beta_p > 0$, and $\Lambda_p(-\mu_p^2)$'s are non-zero. Moreover, we observe that

$$\gamma_p > 0. \tag{A.7}$$

The Lagrange anchor, being defined by the characteristic polynomial $V(\gamma;z)$ (A.1), has the form

$$V(\gamma;z) = \sum_{p=1}^{n} \frac{1}{\beta_p \Lambda_p^2(-\mu_p^2)} \Lambda_p\left(\frac{\partial^2}{\partial x^\mu \partial x_\mu}\right). \tag{A.8}$$

This Lagrange anchor is non-canonical because the coefficient at the highest power of the primary operator is positive.